\documentclass[structabstract]{aa}  
\usepackage{graphicx}
\usepackage{txfonts}

\usepackage{natbib}
\bibpunct{(}{)}{;}{a}{}{,} 

\newcommand{\tg}[1]{{\tt{#1}}}
\newcommand{\ind}[1]{_{\mathrm{#1}}}

\begin{document}
  \title{Possible detection of phase changes from the non-transiting planet HD 46375b by CoRoT\thanks{The CoRoT space mission, launched on 2006 December 27, was developed and is operated by the CNES, with participation of the Science Programs of ESA, ESA's RSSD, Austria, Belgium, Brazil, Germany and Spain.}
}
\author{P.~Gaulme\inst{1}
          \and M.~Vannier\inst{2}
          \and T.~Guillot\inst{3}
          \and B.~Mosser\inst{4}
          \and D.~Mary\inst{2}
          \and W.~W.~Weiss\inst{5}
          \and F.-X.~Schmider\inst{2}
          \and S.~Bourguignon\inst{3}
          \and H.~J.~Deeg\inst{6,7}
          \and C.~R\'egulo\inst{6,7}
          \and S.~Aigrain\inst{8}
          \and J.~Schneider\inst{9}
          \and H.~Bruntt\inst{4}
          \and S.~Deheuvels\inst{4}
          \and J.-F.~Donati\inst{10}
          \and T.~Appourchaux\inst{1}
         \and M.~Auvergne\inst{4} 
          \and A.~Baglin\inst{4}
          \and F.~Baudin\inst{1}
         \and C.~Catala\inst{4}
          \and E.~Michel\inst{4}
         \and  R.~Samadi\inst{4} 
          }
         
\offprints{P. Gaulme}

   \institute{$^1$ Institut d'Astrophysique Spatiale, UMR 8617, Universit\'e Paris Sud, 91405 ORSAY Cedex  \\
$^2$  Laboratoire Fizeau, Universit\'e de Nice, CNRS-Observatoire de la C\^ote d'Azur, 06108 Nice Cedex 2, France\\ 
$^3$  Laboratoire Cassiop\'ee, Universit\'e de Nice, CNRS-Observatoire de la C\^ote d'Azur, 06304 Nice Cedex 4, France\\
$^4$ LESIA, UMR 8109, Observatoire de Paris, 92195 Meudon Cedex, France\\
$^5$ University of Vienna, Inst of Astronomy, T\"urkenschanzstr. 17, AT 1180 Vienna, Austria\\ 
$^{6,7}$ Instituto de Astrof\'isica de Canarias and Universidad de La Laguna, 38205, La Laguna, Tenerife, Spain\\
$^8$ Oxford Astrophysics, University of Oxford, Oxford OX1 3RH, UK\\
$^{9}$ LUTH, UMR 8102, Observatoire de Paris, 92195 Meudon Cedex, France\\
$^{10}$ LATT-UMR 5572, CNRS and University P. Sabatier, 14 Av. E. Belin, 31400 Toulouse, France\\
              \email{Patrick.Gaulme@ias.u-psud.fr}
        }

\titlerunning{reflected light from HD 46375b}
\authorrunning {Gaulme et al.}
\abstract
 {The present work deals with the detection of phase changes in an exoplanetary system. HD~46375 is a solar analog known to host a non-transiting Saturn-mass exoplanet with a 3.0236 day period. It was observed by the CoRoT satellite for 34 days during the fall of 2008.} 
{We attempt to identify at optical wavelengths, the changing phases of the planet as it orbits its star. We then try to improve the star model by means of a seismic analysis of the same light curve and the use of ground-based spectropolarimetric observations.}
{The data analysis relies on the Fourier spectrum and the folding of the time series. }
{We find evidence of a sinusoidal signal compatible in terms of both amplitude and phase with light reflected by the planet. Its relative amplitude is $\Delta F\ind{p}/F\ind{\star} = [13.0, 26.8]$ ppm, implying an albedo $A=[0.16, 0.33]$ or a dayside visible brightness temperature $T\ind{b} \simeq [1880,2030]$~K by assuming a radius $R=1.1 R\ind{Jup}$ and an inclination $i=45^\circ$. Its orbital phase differs from that of the radial-velocity signal by at most $2\,\sigma\ind{RV}$. However, the tiny planetary signal is strongly blended by another signal, which we attribute to a telluric signal with a 1 day period. We show that this signal is suppressed, but not eliminated, when using the time series for HD~46179 from the same CoRoT run as a reference.}
{This detection of reflected light from a non-transiting planet should be confirmable with a longer CoRoT observation of the same field. In any case, it demonstrates that non-transiting planets can be characterized using ultra-precise photometric lightcurves with present-day observations by CoRoT and Kepler. The combined detection of solar-type oscillations on the same targets (Gaulme et al. 2010a) highlights the overlap between exoplanetary science and asteroseismology and shows the high potential of a mission such as Plato.}

\maketitle

\section{Introduction}

The star HD~46375 is a relatively bright ($m\ind{V}=7.89$, $M\ind{bol} = 5.04$) K0 star, known to host a Saturn-mass planet  ($M\ind{p} \sin i = 0.226(19)\ M\ind{Jup}$) orbiting in 3.023573(65) days, at $0.0398(23)$ AU, on a quasi-circular orbit $e = 0.063(26)$ (\citealt{Marcy_2000}, \citealt{Butler_2006}). The planet does not transit its host star \citep{Henry_2000}. HD~46375 is the brightest star with a known close-in planet in the CoRoT accessible field of view. As such, it was targeted by the CoRoT additional program and observed in a CCD normally dedicated to the asteroseismology program \citep[][]{Baglin_2002}, to obtain an ultra-precise photometric lightcurve and detect or place upper limits on the brightness of the planet \citep[][]{Guillot_Vannier_2003}. 

The detection of reflected light from an exoplanetary atmosphere is very interesting for several reasons: the amplitude of the modulation helps us to constrain the atmospheric properties, and in particular the planetary albedo. This has direct consequences for the calculation of the thermal evolution of these planets and therefore their structure. Departures from a sinusoidal shape may be caused by non-isotropic scattering and betray the presence of small particles in the atmospheres \citep[e.g.,][]{Seager_2000}. Finally, the asymmetry of the modulation or a phase shift may be due to horizontal variations in the cloud coverage and/or atmospheric temperatures and trace the atmospheric circulation \citep[e.g.][]{Showman_Guillot_2002}.  

In the optical domain, this detection is extremely challenging because of the faintness of the signal, of the order of $10^{-4}$ (100~ppm) in the most favorable cases and was achieved in only a few cases. \citet{Sing_Morales_2009} detected the secondary transit of OGLE-TR-56b in the z'-band, a case for which the planet is so hot that thermal emission from the planet contributes to the $363\pm91$ ppm signal. Using CoRoT data, \citet{Snellen_2009} identified the changing phases of the planet CoRoT-1b (with an amplitude $\sim 130$ ppm) and its secondary transit was independently measured by \citet{Alonso_2009a}. Secondary transits of CoRoT-2b (a $60\pm20$ ppm signal) were detected by \citet{Alonso_2009b}. With Kepler data, \citet{Borucki_2009} measured the secondary eclipse of HAT-P-7b {\tg($130\pm 11$ ppm)}. With HD~46375, we have the opportunity to the detect planetary phase changes of a non-transiting exoplanet. Such a kind detection was achieved in the infrared with the Spitzer Space Telescope for $\upsilon$~And \citep{Harrington_2006} (amplitude $\sim 2900$ ppm) but it would be the first time for optical wavelengths. 

In a separate paper (Gaulme et al. 2010a, A\&A, submitted), we used spectroscopic data and seismic analysis of the CoRoT lightcurves to constrain the fundamental parameters of HD~46375. The star may be considered as a solar analog, of mass $0.97\pm0.05\ M_{\sun}$, age $2.6\pm0.8$ Gyr, metallicity [Fe/H] $= 0.39\pm0.06$ dex, and temperature $T\ind{eff} = 5300\pm60$ K. The stellar rotation is expected to be in the range $[37, 51]$ days. The spectrometric and asteroseismic results provide an estimate of the inclination of the stellar rotation axis $i_*=50\pm 18^\circ$ and the planetary mass $M\ind{p}\sin i = 0.234\pm 0.008\ M\ind{Jup}$. By assuming that planet orbits in the stellar equatorial plane \citep[see however][]{Hebrard_2010}, yields a true planetary mass $M\ind{p} = 0.30^{+0.14}_{-0.05}\ M\ind{Jup}$.

In this paper, we use CoRoT observations to search for light from the planetary companion. We present successively the expected performance (Sect. 2), then, the data analysis in the Fourier and the time domain to ensure an ambiguous detection of the planetary light (Sect. 3). In Sect. 4, we identify a spurious systematic signal in the CoRoT light curve and propose an approximate correction. We conclude that we achieve a possible detection of the planetary contribution to the light curve, which would  need further and longer observations to strengthen the result (Sect. 6).

\section{Expected signal}
We neglected the eccentricity of the orbit and assumed that the starlight was reflected isotropically over the ``dayside'' of the planet (Lambert sphere). The peak-to-peak amplitude of the photometric fluctuations caused by the planet is
\begin{equation}
\frac{\Delta F\ind{p}(t)}{F_\star(t)}\ =\ \frac{2A}{3}\ \left(\frac{R}{a}\right)^2  \sin{i},
\label{eq_dphi_sur_phi}
\end{equation}
where $F\ind{p}$ and $F_\star(t)$ are the planetary 	and stellar fluxes, $A$ and $R$ the planetary Bond albedo and radius, $a$ the orbital distance, and $i$ the orbital inclination \citep[e.g.][]{charbonneau_1999}. There are therefore three degenerate parameters: the albedo, the orbital inclination, and the planetary radius. 

The planetary radius may be estimated from evolution calculations that are parameterized to reproduce the measured radii of transiting exoplanets \citep{Guillot2008}. The result depends mostly on the unknown core mass: for an assumed planetary mass of $0.3\rm\,M_{Jup}$ and a dissipation of 1\% of the incoming stellar light at the planet's center, we derive planetary radii ranging from $0.8\rm\,R_{Jup}$ for a $60\rm\,M_\oplus$ dense core to $1.6\rm\,R_{Jup}$ for a planet with no core. 

By assuming that the orbital plane is tilted by $i = 45^\circ$, we found the radius $R\ind{p}=1.1\,R_{Jup}$ and given the large assumed uncertainty in the planetary albedo, $A=0.05$ to $0.5$, the relative amplitude  $\Delta F\ind{p}/F_\star$ of the expected signal is in the range $ [0.8, 61]$~ppm. Such relative fluctuations of intensity have to be compared with the photon noise. In a single channel, the amount of photoelectrons collected by the CCD is about $N_\star=1.3\times10^{9}\times10^{-0.4 m_V}\,{\rm s}^{-1}$. Hence, for a  $m\ind{V}=7.9$ star, the relative photon noise is $1/\sqrt{N_\star}= 10^{-3}$ per second. If individual measurements are binned in 6 samples per orbital period, the noise level reaches 1.5 ppm for each orbital sample, for a time coverage of 11 orbits (i.e., 33 days). Hence, the expected signal-to-noise ratio of the planetary signal is $[0.5, 40]$ (peak-to-peak).

Therefore, in terms of photon noise, the precision of CoRoT should be sufficient, except for a combination of the lower bounds of both inclination and albedo. Note that we have not investigated the case where the reflection is anisotropic {\tg(e.g. \citealt{Guillot_Vannier_2003}).} A hypothetical non-strictly sinusoidal planetary signal would include harmonics of the orbital frequency. Since their values are {\it a priori} unknown, and given that the stellar signal itself displays components of relatively high and variable amplitudes at frequencies higher than the orbital one, we estimate that the planetary effect will only be detectable in terms of its main orbital frequency (Eq. 1).

\begin{figure}[t]
\includegraphics[width=8.5cm]{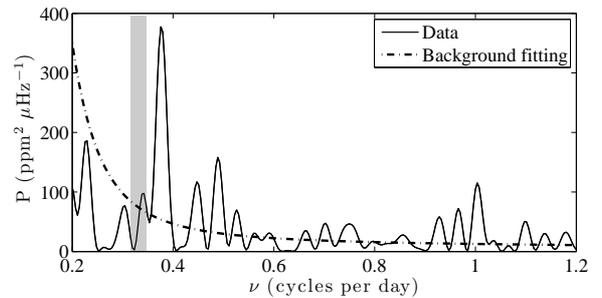}
\caption[]{Power density spectrum calculated with DFT, with an oversampling factor of 8. The dot-dashed curve indicates the mean stellar activity. The grey region indicates the range in which we expect to detect the planetary signal.
}
\label{DSP_avant}
\end{figure}

\section{Ambiguous result from data analyses}
\subsection{No clear planetary signature in the Fourier spectrum}
\label{fourier}
The CoRoT light curve (Fig. \ref{vraie_lightcurve}) does not exhibit signs of high activity. Only relative variations smaller than 400 ppm are observed on typical timescales of 10 days. On the one hand, the lack of activity aids the identification of the planetary signal embedded in the lightcurve. On the other hand, the low activity coupled with the long stellar rotation period ensures that it is impossible for the spot analysis to be performed in a similar way to \citet{Mosser_2009_taches}. Therefore the stellar rotation estimate cannot be refined, complicating the disentangling of the planetary from the stellar signal.

We first search for signatures of the planetary changing phases in the Fourier spectrum. From \citet{Butler_2006}, the planetary orbital period is $3.023573 \pm 0.000065$ day, which corresponds to the frequency $0.330735 \pm 0.000007$ cycles per day (c/d). The frequency resolution of a 34-day run is 0.029 c/d. Hence, the planetary signal should appear as a peak in the frequency range $[0.32, 0.35]$ c/d. In Fig. \ref{DSP_avant}, we present the power density spectrum calculated using the discrete Fourier transform (DFT). The background, whose amplitude is the result of the stellar activity, was reproduced by fitting a sum of 2 semi-Lorentzian functions to the whole power spectrum (Gaulme et al. 2010a).

No excess power appears in the expected frequency range. Moreover, we note the peak at 0.38 c/d, whose origin is not explained by typical systematic signals of the CoRoT, such as the orbital period (13.97 c/d), the Earth rotation period (visible around 1 c/d), or a combination of them. Therefore, the Fourier analysis is inconclusive and a direct analysis of the time series has to be performed, by taking into account the information about the planetary orbital phase.

\subsection{A sine trend in the folded time series}

\begin{figure}[t]
\includegraphics[width=8.5cm]{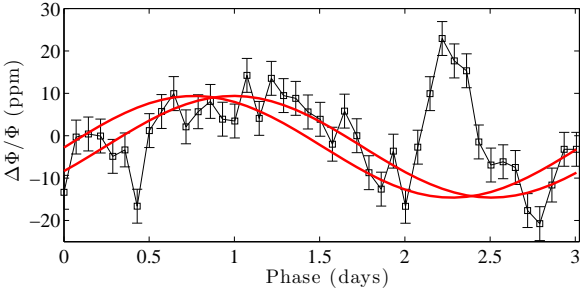}
\caption[]{Folded time series over 3.023573 days. The 2 red sine curves represent the minimum and maximum limits of the expected signal according to the ephemeris of Butler et al. 2006.}
\label{folding_375_et_179}
\end{figure}
\begin{figure}[t]
\includegraphics[width=8.5cm]{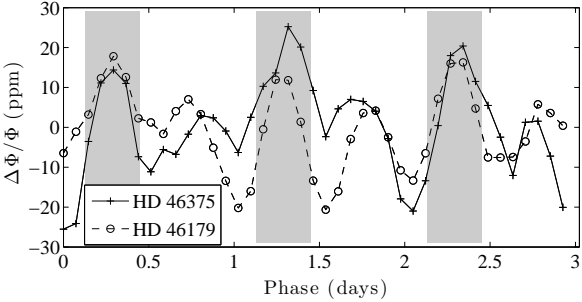}
\caption[]{Folding of the HD~46375 and HD~46179 time series over 3 days. The grey zones correspond to region where the correlation between both light curves is maximum.}
\label{folding_3days}
\end{figure}

The alternative way to search for the planetary signal, for which both phase and period are known, consists of folding the time series over the planetary period. To increase the signal-to-noise ratio, we first rebinned the data over 2 CoRoT orbital periods, i.e. 6184 s. Then, the low frequency trend, consisting of the moving average weighted by a triangular window of 7-day FWHM, is subtracted from the time series. We note that we checked this step does not affect the planetary period. The rebinned and flattened time series is folded over the planetary period. If the noise is dominated by photon noise, the standard deviation of the time series is 176 ppm (Fig. \ref{vraie_lightcurve}) for $92\,140$ data measurements gives for the folded time series (43 bins) has a 3.9 ppm noise level.

The analysis of the folded time series over the planetary orbit detects two main trends (Fig. \ref{folding_375_et_179}). Firstly, a sine trend with a peak-to-peak amplitude of about 20 ppm appears to be in phase with the expected signal. Secondly, a sharp excess power is visible in the range $t =[2.2, 2.5]$ days, which has an amplitude of about 35 ppm. We must explain the origin of the tall sharp peak, to before we can be certain that the sine trend is not an artefact.

\section{Evidence of a signal compatible with the planet}
\subsection{A 24-h periodic systematic signal}
The peculiarity and difficulty of HD~46375b is that its orbital period is close to an integer number of days, which makes its detection very sensitive to telluric systematic signals. Contaminating signals with 1-day periods were shown to be present in the CoRoT exoplanetary lightcurves \citep{Mazeh_2009}.

\begin{figure}[t]
\includegraphics[width=8.5cm]{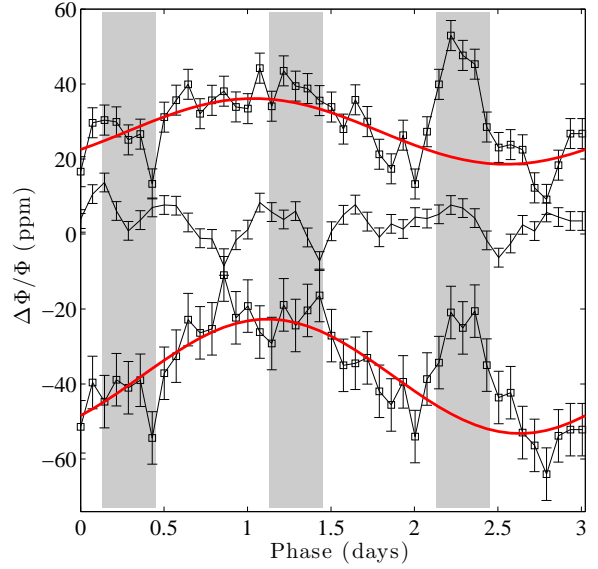}
\caption[]{Top: folding of HD~46375 time series over the planetary period. Middle: folding of HD~46179 over the same period. Bottom: folding of the time series made of the difference between HD~46375 and HD~46179 data. The red sine curves indicate the data fitting by excluding the regions indicated in Fig. \ref{folding_3days}.}
\label{folding_375_moins_179}
\end{figure}

In our case, among the 9 stars followed up by CoRoT in the same observation run in the seismic field, only HD~46179 ($m\ind{Hip} = 6.68$) exhibits less activity than HD~46375, with variations below 165 ppm after removing a linear trend. We processed the light curve in the same way as HD~46375, and then folded both time series exactly over 3 days (Fig \ref{folding_3days}). The 2 curves appear to be strongly correlated with a clear 24-h period. Therefore, a common systematic signal is present in both lightcurves. This feature is probably related to the CoRoT orbit, and may be caused by e.g. flux variations generated by the entrance/exit in the night side of the Earth \citep{Auvergne_2009}. We note that only HD~46375 exhibits a sine trend when folded about the planetary period, which indicated that the sine trend is characteristic of HD~46375 (Fig. \ref{folding_375_moins_179}).

\begin{figure}[t]
\includegraphics[width=8.5cm]{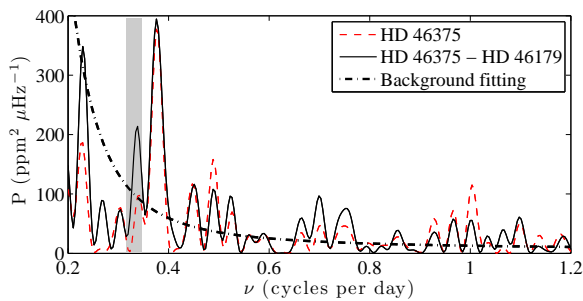}
\caption[]{Power density spectra of the time series corresponding to the difference between HD~46375 and HD~46179 lightcurves, compared to HD~46375.}
\label{DSP_apres}
\end{figure}

\subsection{Cleaning of the time series}

We first tried to clean the lightcurves using the SYS-REM algorithm (Guterman \& Mazeh, private comm.). However, this did not lead to a reduction in the standard deviation at low frequencies. We therefore adopted a simple cleaning method, by considering the HD~46179 data as reference. Firstly, we flattened the HD~46179 lightcurve by removing a linear trend, and then subtracted it from the HD~46375 time series. The rebinned and folded time series exhibits a clearer sine trend, while the relative amplitude with respect to the sine amplitude is reduced by about 2  (Fig. \ref{folding_375_moins_179}).

Moreover, the Fourier analysis confirms that while the spectrum amplitude decreases at both 1~c/d and 0.5~c/d, it increases significantly in the frequency range which including both the terrestrial harmonics (1/3~c/d) and HD~46375b orbital period (Fig. \ref{DSP_apres}). However, since the light due to Earth-shine is unlikely to follow a simple path to reach the detector, we cannot reject the hypothesis that the increase in power at 0.33 c/d could be an artefact, introduced by the comparison of two light curves obtained on different parts of the detector.

\subsection{A signal compatible with HD~46375b}
The variability in the 1-day period spurious signal in both time and space prevents us from properly modeling it and cleaning the data. From Fig. \ref{folding_3days}, the intensity of this spurious signal reaches a maximum in  0.3-day long intervals close to phases $[0.3, 1.3, 2.3]$ days. We therefore fitted the folded time series by excluding these 3 ranges, to minimise the influence of the systematics on the sine fitting. The fitting function is a simple sine, whose phase $\phi$, amplitude $\Delta  F\ind{p}/F\ind{\star}$, and offset are free. The harmonics of the planetary orbit are left aside, because of the low signal-to-noise level. 

We first fitted the raw, then the ``roughly cleaned'' light curves. The least squares fitting (reduced $\chi^2 = 1.75$) measured a relative peak-to-valley amplitude $\Delta F\ind{p}/F\ind{\star} = 13.0\pm5.0$ ppm and a phase shift of $19\pm9^\circ$ with respect to the ephemeris. A phase shift of $19^\circ$ corresponds to 1.4-$\sigma$ with the \citet{Butler_2006} error bar on the elongation date, which is reasonable since their value lies on 50 measurements, all performed before 2006. The fitting of the ``cleaned'' data allows us to improve the fitting ($\chi^2 = 0.87$) and to re-estimate the sine amplitude $\Delta F\ind{p}/F\ind{\star}=26.8\pm3.0$ ppm, with a phase shift  $\phi = 29\pm 6^\circ$, i.e. corresponding to a 2-$\sigma$ error relative to the predicted ephemeris. Moreover, the amplitude of the peak corresponding to the planetary period is significantly larger in the density spectrum of the cleaned data (Fig. \ref{DSP_apres}). 

After fitting the folded time series by excluding three regions, we identified a signal compatible with the planet. Assuming that $i=45^\circ$, $R=1.1\rm\,R_{Jup}$, and $\Delta F\ind{p}/F\ind{\star} = [13.0, 26.8]$~ppm, our inferred albedo is between 0.16 and 0.33, which is higher than expected on the basis of theoretical models, but within the range of actually derived albedos \citep[see][]{Cowan_Agol_2010}. Slightly larger radii and inclinations, of course, yield lower albedos consistent with theoretical expectations \citep[see][]{Sudarsky_2003}. Alternatively, the observed flux may be produced by direct thermal emission. The brightness temperatures can be calculated from the ratio of the black body emissions of the planet with respect to the star, weighted by the CoRoT transmission function \citep{Auvergne_2009}. We found that the minimum temperature is at least $[1880, 2030]$~K for $\Delta F\ind{p}/F\ind{\star} = [13.0, 26.8]$~ppm, by assuming that the night side has a negligible emission and the planet is seen almost edge-on. This is unlikely, because the ratio of dayside optical brightness temperature to equilibrium temperature would be at least between $1.6$ and $1.8$ for HD~46375b compared to measured values of between $1.2$ and $1.45$ for OGLE-TR-56b, CoRoT-1b, and CoRoT-2b.

\section{Discussion and conclusion}

Our analysis of  observation data presented here has been the first attempt to directly detect the stellar light reflected by a non-transiting planet in the visible domain. In parallel to the present work, we have performed a seismic analysis on the same data set. Coupled with ground-based spectropolarimetric observations, obtained with the NARVAL spectropolarimeter at the Pic du Midi Observatory, we have been able to strongly refine the stellar modelling (Gaulme et al. 2010a). 

We used two approaches to detect the planetary contribution to the CoRoT light curve. On the one hand, the orbital signal was only barely detected by our Fourier analysis. On the other hand, the time series analysis, consisting of rebinning and folding the light curve over the planetary orbital period, has allowed us to detect a signal compatible with the planetary orbital signature. The folded light curve appears to be composed of two mains signals: a sine trend, in phase with ephemeris, and a high excess of intensity, which has no a priori explanation.

We compared our data with the time series of HD~46179, which is the only star among those of the same seismic CoRoT run to exhibit a low activity, which is actually lower than that of HD~46375. The 3-day folding of both time series enabled us to detect a common 1-day period systematic signal, with $\approx 35$ ppm maximum amplitude (Fig. \ref{folding_3days}). We reduced by a factor 2 the relative amplitude of the spurious peak with respect to the sine trend, by subtracting the filtered time series of HD~46179 from the HD~46375 time series. 

We illustrated that the data is affected by the presence of a systematic signal presumedly of telluric origin, which is strongly variable in both time and space. It appears necessary to study the photometric fluctuations at low frequency in the whole 3-year CoRoT set data especially in the seismic field, to refine the modeling of spurious signals. This objective goes far beyond the aims of the present work. However, the only way of determining whether that the sine trend is of planetary origin consists of observing the HD~46375 system for a duration long enough to disentangle the planetary orbital period from 3 days, i.e. a duration longer than $\sim 100$ days, as achievable with a CoRoT long run.

\bibliographystyle{aa} 
\bibliography{bibi} 

\begin{thebibliography}{21}
\expandafter\ifx\csname natexlab\endcsname\relax\def\natexlab#1{#1}\fi

\bibitem[{{Alonso} {et~al.}(2009{\natexlab{a}}){Alonso}, {Alapini}, {Aigrain},
  {Auvergne}, {Baglin}, {Barbieri}, {Barge}, {Bonomo}, {Bord{\'e}}, {Bouchy},
  {Chaintreuil}, {de La Reza}, {Deeg}, {Deleuil}, {Dvorak}, {Erikson},
  {Fridlund}, {de Oliveira Fialho}, {Gondoin}, {Guillot}, {Hatzes}, {Jorda},
  {Lammer}, {L{\'e}ger}, {Llebaria}, {Magain}, {Mazeh}, {Moutou}, {Ollivier},
  {P{\"a}tzold}, {Pont}, {Queloz}, {Rauer}, {Rouan}, {Schneider}, \&
  {Wuchterl}}]{Alonso_2009b}
{Alonso}, R., {Alapini}, A., {Aigrain}, S., {et~al.} 2009{\natexlab{a}}, \aap,
  506, 353

\bibitem[{{Alonso} {et~al.}(2009{\natexlab{b}}){Alonso}, {Guillot}, {Mazeh},
  {Aigrain}, {Alapini}, {Barge}, {Hatzes}, \& {Pont}}]{Alonso_2009a}
{Alonso}, R., {Guillot}, T., {Mazeh}, T., {et~al.} 2009{\natexlab{b}}, \aap,
  501, L23

\bibitem[{{Auvergne} {et~al.}(2009){Auvergne}, {Bodin}, {Boisnard}, {Buey},
  {Chaintreuil}, {Epstein}, {Jouret}, {Lam-Trong}, {Levacher}, {Magnan},
  {Perez}, {Plasson}, {Plesseria}, {Peter}, {Steller}, {Tiph{\`e}ne}, {Baglin},
  {Agogu{\'e}}, {Appourchaux}, {Barbet}, {Beaufort}, {Bellenger}, {Berlin},
  {Bernardi}, {Blouin}, {Boumier}, {Bonneau}, {Briet}, {Butler}, {Cautain},
  {Chiavassa}, {Costes}, {Cuvilho}, {Cunha-Parro}, {de Oliveira Fialho},
  {Decaudin}, {Defise}, {Djalal}, {Docclo}, {Drummond}, {Dupuis}, {Exil},
  {Faur{\'e}}, {Gaboriaud}, {Gamet}, {Gavalda}, {Grolleau}, {Gueguen},
  {Guivarc'h}, {Guterman}, {Hasiba}, {Huntzinger}, {Hustaix}, {Imbert},
  {Jeanville}, {Johlander}, {Jorda}, {Journoud}, {Karioty}, {Kerjean},
  {Lafond}, {Lapeyrere}, {Landiech}, {Larqu{\'e}}, {Laudet}, {Le Merrer},
  {Leporati}, {Leruyet}, {Levieuge}, {Llebaria}, {Martin}, {Mazy}, {Mesnager},
  {Michel}, {Moalic}, {Monjoin}, {Naudet}, {Neukirchner}, {Nguyen-Kim},
  {Ollivier}, {Orcesi}, {Ottacher}, {Oulali}, {Parisot}, {Perruchot},
  {Piacentino}, {Pinheiro da Silva}, {Platzer}, {Pontet}, {Pradines},
  {Quentin}, {Rohbeck}, {Rolland}, {Rollenhagen}, {Romagnan}, {Russ}, {Samadi},
  {Schmidt}, {Schwartz}, {Sebbag}, {Smit}, {Sunter}, {Tello}, {Toulouse},
  {Ulmer}, {Vandermarcq}, {Vergnault}, {Wallner}, {Waultier}, \&
  {Zanatta}}]{Auvergne_2009}
{Auvergne}, M., {Bodin}, P., {Boisnard}, L., {et~al.} 2009, \aap, 506, 411

\bibitem[{{Baglin} {et~al.}(2002){Baglin}, {Auvergne}, {Barge}, {Buey},
  {Catala}, {Michel}, {Weiss}, \& {COROT Team}}]{Baglin_2002}
{Baglin}, A., {Auvergne}, M., {Barge}, P., {et~al.} 2002, in ESA Special
  Publication, Vol. 485, Stellar Structure and Habitable Planet Finding, ed.
  {B.~Battrick, F.~Favata, I.~W.~Roxburgh, \& D.~Galadi}, 17--24

\bibitem[{{Borucki} {et~al.}(2009){Borucki}, {Koch}, {Jenkins}, {Sasselov},
  {Gilliland}, {Batalha}, {Latham}, {Caldwell}, {Basri}, {Brown},
  {Christensen-Dalsgaard}, {Cochran}, {DeVore}, {Dunham}, {Dupree}, {Gautier},
  {Geary}, {Gould}, {Howell}, {Kjeldsen}, {Lissauer}, {Marcy}, {Meibom},
  {Morrison}, \& {Tarter}}]{Borucki_2009}
{Borucki}, W.~J., {Koch}, D., {Jenkins}, J., {et~al.} 2009, Science, 325, 709

\bibitem[{{Butler} {et~al.}(2006){Butler}, {Wright}, {Marcy}, {Fischer},
  {Vogt}, {Tinney}, {Jones}, {Carter}, {Johnson}, {McCarthy}, \&
  {Penny}}]{Butler_2006}
{Butler}, R.~P., {Wright}, J.~T., {Marcy}, G.~W., {et~al.} 2006, \apj, 646, 505

\bibitem[{{Charbonneau} {et~al.}(1999){Charbonneau}, {Noyes}, {Korzennik},
  {Nisenson}, {Jha}, {Vogt}, \& {Kibrick}}]{charbonneau_1999}
{Charbonneau}, D., {Noyes}, R.~W., {Korzennik}, S.~G., {et~al.} 1999, \apjl,
  522, L145

\bibitem[{{Cowan} \& {Agol}(2010)}]{Cowan_Agol_2010}
{Cowan}, N.~B. \& {Agol}, E. 2010, ArXiv e-prints

\bibitem[{{Guillot}(2008)}]{Guillot2008}
{Guillot}, T. 2008, Physica Scripta Volume T, 130, 014023

\bibitem[{{Guillot} \& {Vannier}(2003)}]{Guillot_Vannier_2003}
{Guillot}, T. \& {Vannier}, M. 2003, in EAS Publications Series, Vol.~8, EAS
  Publications Series, ed. {C.~Aime \& R.~Soummer}, 25--35

\bibitem[{{Harrington} {et~al.}(2006){Harrington}, {Hansen}, {Luszcz},
  {Seager}, {Deming}, {Menou}, {Cho}, \& {Richardson}}]{Harrington_2006}
{Harrington}, J., {Hansen}, B.~M., {Luszcz}, S.~H., {et~al.} 2006, Science,
  314, 623

\bibitem[{{Hebrard} {et~al.}(2010){Hebrard}, {Desert}, {Diaz}, {Boisse},
  {Bouchy}, {Lecavelier des Etangs}, {Moutou}, {Ehrenreich}, {Arnold},
  {Bonfils}, {Delfosse}, {Desort}, {Eggenberger}, {Forveille}, {Gregorio},
  {Lagrange}, {Lovis}, {Pepe}, {Perrier}, {Pont}, {Queloz}, {Santerne},
  {Santos}, {Segransan}, {Sing}, {Udry}, \& {Vidal-Madjar}}]{Hebrard_2010}
{Hebrard}, G., {Desert}, J., {Diaz}, R.~F., {et~al.} 2010, ArXiv e-prints

\bibitem[{{Henry}(2000)}]{Henry_2000}
{Henry}, G.~W. 2000, \apjl, 536, L47

\bibitem[{{Marcy} {et~al.}(2000){Marcy}, {Butler}, \& {Vogt}}]{Marcy_2000}
{Marcy}, G.~W., {Butler}, R.~P., \& {Vogt}, S.~S. 2000, \apjl, 536, L43

\bibitem[{{Mazeh} {et~al.}(2009){Mazeh}, {Guterman}, {Aigrain}, {Zucker},
  {Grinberg}, {Alapini}, {Alonso}, {Auvergne}, {Barbieri}, {Barge},
  {Bord{\'e}}, {Bouchy}, {Deeg}, {de La Reza}, {Deleuil}, {Dvorak}, {Erikson},
  {Fridlund}, {Gondoin}, {Jorda}, {Lammer}, {L{\'e}ger}, {Llebaria}, {Magain},
  {Moutou}, {Ollivier}, {P{\"a}tzold}, {Pont}, {Queloz}, {Rauer}, {Rouan},
  {Sabo}, {Schneider}, \& {Wuchterl}}]{Mazeh_2009}
{Mazeh}, T., {Guterman}, P., {Aigrain}, S., {et~al.} 2009, \aap, 506, 431

\bibitem[{{Mosser} {et~al.}(2009){Mosser}, {Baudin}, {Lanza}, {Hulot},
  {Catala}, {Baglin}, \& {Auvergne}}]{Mosser_2009_taches}
{Mosser}, B., {Baudin}, F., {Lanza}, A.~F., {et~al.} 2009, \aap, 506, 245

\bibitem[{{Seager} {et~al.}(2000){Seager}, {Whitney}, \&
  {Sasselov}}]{Seager_2000}
{Seager}, S., {Whitney}, B.~A., \& {Sasselov}, D.~D. 2000, \apj, 540, 504

\bibitem[{{Showman} \& {Guillot}(2002)}]{Showman_Guillot_2002}
{Showman}, A.~P. \& {Guillot}, T. 2002, \aap, 385, 166

\bibitem[{{Sing} \& {L{\'o}pez-Morales}(2009)}]{Sing_Morales_2009}
{Sing}, D.~K. \& {L{\'o}pez-Morales}, M. 2009, \aap, 493, L31

\bibitem[{{Snellen} {et~al.}(2009){Snellen}, {de Mooij}, \&
  {Albrecht}}]{Snellen_2009}
{Snellen}, I.~A.~G., {de Mooij}, E.~J.~W., \& {Albrecht}, S. 2009, \nat, 459,
  543

\bibitem[{{Sudarsky} {et~al.}(2003){Sudarsky}, {Burrows}, \&
  {Hubeny}}]{Sudarsky_2003}
{Sudarsky}, D., {Burrows}, A., \& {Hubeny}, I. 2003, \apj, 588, 1121

\end{thebibliography}

\end{document}